**A computational account of the development and evolution of psychotic symptoms**

Albert Powers[1*], Philip Angelos[1], Alexandria Bond[1], Emily Farina[1], Carolyn Fredericks[1], Jay Gandhi[1], Maximillian Greenwald[1], Gabriela Hernandez-Busot[1], Gabriel Hosein[1], Megan Kelley[1], Catalina Mourgues[1], William Palmer[2], Julia Rodriguez-Sanchez[3], Rashina Seabury[1], Silmilly Toribio[1], Raina Vin[1], Jeremy Weleff[1], David Benrimoh[4]

1. Yale University School of Medicine and the Connecticut Mental Health Center, New Haven, CT, USA
2. Yale University Department of Psychology, New Haven, CT USA
3. Centre for Medical Image Computing, University College London
4. Department of Psychiatry, McGill University, Montreal, Canada

*Correspondence should be addressed to:

Albert R. Powers, M.D., Ph.D.
The Connecticut Mental Health Center, Rm. S109
34 Park Street
New Haven, CT 06519
albert.powers@yale.edu
203.974.7329

*Key words: computational psychiatry; hallucinations; perception; psychosis; nosology*

Words in abstract: 250

Words in main body of manuscript: 3887

Figures: 1

**Abstract**

The mechanisms of psychotic symptoms like hallucinations and delusions are often investigated in fully-formed illness, well after symptoms emerge. These investigations have yielded key insights, but are not well-positioned to reveal the dynamic forces underlying symptom formation itself. Understanding symptom development over time would allow us to identify steps in the pathophysiological process leading to psychosis, shifting the focus of psychiatric intervention from symptom alleviation to prevention.

We propose a model for understanding the emergence of psychotic symptoms within the context of an adaptive, developing neural system. We will make the case for a pathophysiological process that begins with cortical hyperexcitability and bottom-up noise transmission, which engenders inappropriate belief formation via aberrant prediction error signaling. We will argue that this bottom-up noise drives learning about the (im)precision of new incoming sensory information because of diminished signal-to-noise ratio, causing an adaptive relative over-reliance on prior beliefs. This over-reliance on priors predisposes to hallucinations and covaries with hallucination severity. An over-reliance on priors may also lead to increased conviction in the beliefs generated by bottom-up noise and drive movement toward conversion to psychosis. We will identify predictions of our model at each stage, examine evidence to support or refute those predictions, and propose experiments that could falsify or help select between alternative elements of the overall model.

Nesting computational abnormalities within longitudinal development allows us to account for hidden dynamics among the mechanisms driving symptom formation and to view established symptomatology as a point of equilibrium among competing biological forces.

**Introduction**

Modern medicine depends on a basic knowledge of pathophysiology. By understanding the bodily states that drive normal functioning, we can begin to understand how these are disrupted over time to produce disease. That understanding then drives identification of new ways to interrupt and reverse the disease process for prevention and treatment (1,2).

Calls for psychiatry to take a similar approach have grown (3,4), but difficulties in identifying biomarkers for established psychiatric diagnoses have slowed progress. Renewed hope in these efforts has come with the advent of theory-based computational approaches to understanding normative brain function and how its derangement might produce psychiatric disease (5–7). Theory-based approaches, through the use of explicitly-defined mathematical models, have the power to connect from the microscopic level of conserved internal processes (8,9) to the macroscopic level of observable behavior and symptoms (10,11). This comprehensive framework not only bridges the gap between internal mechanisms and external manifestations of psychiatric diseases but also enables the testing of a broad spectrum of hypotheses regarding unobserved (i.e., latent) states that underlie symptomatology and disease progression.Recent studies based on these generative modeling approaches have succeeded in identifying information processing abnormalities that drive abnormal behavior in participants with symptoms as diverse as hallucinations (12,13), anxiety (14,15), and binge eating (16), among many others.

Because of their ability to estimate latent states driving behavior relevant to symptomatology, generative modeling approaches have the potential to track the relationships between these latent states and symptoms over time (13). This brings our knowledge of the causal chain driving symptoms to the step just prior to symptom onset and also already makes these approaches immensely powerful as tools for prognosis: if latent state estimates are found to change prior to symptom worsening, they may serve as early warning signs and would allow for directed early intervention before a crisis arises.

But, perhaps because it is difficult to obtain data pertaining to initial symptom development, these approaches have not yet been used to their full potential in identifying the causal chain(s) of latent states that ultimately lead to symptom development. To draw a parallel with cardiology, identification of latent states leading directly to psychiatric symptom development is akin to identifying why a patient has angina–namely, they have decreased blood supply to their cardiac muscle that then leads to chest pain. This is an important realization and can lead directly to intervention by alleviating the causal narrowing of the coronary arteries. But the real impact comes from understanding the precursor processes like atherosclerosis that lead to narrowed arteries in the first place. This insight then makes primary prevention possible via statins, antihypertensives, and lifestyle modifications. In understanding psychiatric symptom development, the real benefit in the form of prevention and early intervention will come from identification of processes that lead to latent states that lead directly to symptom development.

In this invited review, we outline a model for how derangements in information processing that appear to underlie psychotic symptom expression may arise. Our hope is that, by identifying the potential links in such a causal chain, we might test and refine this model, ultimately expanding the scope of psychiatric practice from symptom alleviation to prevention.

**Predictive processing and psychotic symptoms**

Historically grouped into the positive symptoms of psychosis, delusions and hallucinations both involve alterations in a person's sense of reality. The presence of at least one of these primary symptoms of psychosis is required for clinical diagnosis of a psychotic disorder (17). Although they often co-occur in psychotic illness, recent evidence suggests that delusions and hallucinations may emerge from distinct alterations in learning and perceptual inference (18–23).

Based in the predictive processing computational framework (24) and the related Bayesian Brain hypothesis (25), many theories have emerged to describe how individual psychotic symptoms might emerge from alterations in how the brain models its environment and incorporates new information (20,21,26–28). In brief, these ideas describe an efficient and evolutionarily-conserved way the brain might best understand the contents of its environment and act upon it to achieve goals crucial to the organism's survival--namely, it must build an internal model of its environment and use new information from the senses to update that model, depending on the relative reliabilities of the model itself and new sensory information (29–31). Integrating new information into existing beliefs is optimally achieved via Bayesian statistical mechanisms in which beliefs are updated when there is a mismatch between predicted and observed data (i.e., a prediction error), weighted by the relative reliabilities of the prediction and the prediction error. Multiple perceptual and cognitive processes have been shown to approximate Bayes-optimal processes (32). This simple scheme provides a lens through which we might view all functions that involve inference about the environment and formation and updating of new beliefs, including ways normative processing might be altered to produce errors that could produce psychiatric symptoms.

Within this framework, delusions may be thought to arise from inappropriate learning about the world, resulting in beliefs that do not conform to consensual reality (27). Behavioral and imaging findings consistently relate inappropriate prediction error signaling to a propensity toward delusional ideation in psychosis (33), schizotypy (34,35), and in healthy individuals under the influence of the psychotomimetic drug ketamine (36,37). Interestingly, paranoid ideation may relate specifically to enhanced beliefs about environmental volatility (38,39), possibly stemming initially from enhanced prediction error signaling supporting learning about that volatility. It should also be noted that this work specifically relates to delusion formation, rather than maintenance; delusional conviction has been shown to specifically relate to the more top-down jumping-to-conclusions bias (38), which we will revisit in a later section.

Bayesian formulations of perception describe it as an inferential process–coming to a conclusion about the content of the environment using existing beliefs about that environment and new information from the senses. If this is true, it stands to reason that hallucinations (i.e., percepts that do not correspond to sensory evidence) may result from an over-weighting of existing beliefs during that process (18,21,28). This has now been borne out by multiple different experiments using different paradigms across different disease processes and in the general population (12,13,40–45). The consistent finding across these studies is that individuals with hallucinations are more likely to rely on their expectations (or *priors* in Bayesian terms) than on their sensory evidence. It also appears that this susceptibility not only reflects long-term propensity toward hallucinations but tracks with hallucination severity (13,43), supporting the idea that hyper-precise priors are a state of information processing leading directly to symptom expression.

Of course, delusions and hallucinations co-occur in psychotic illness, which leaves us to explain how the same symptoms might co-exist despite purportedly being caused by seemingly opposite information processing abnormalities. In attempting to explain this apparent contradiction, some have rightly appealed to the fact that priors and prediction errors occur at each level of a vast neural processing hierarchy spanning perception and cognition (26,28). Supporting this account, recent work has emphasized differences in low-level sensory and higher-level volatility-based prediction error signaling in psychosis (46,47).

In this piece, we propose a model that accounts for this seeming contradiction by positing that the abnormalities driving the emergence of delusions and hallucinations are causally related over time (**Fig. 1**): an over-reliance on priors that predisposes to hallucinations may be a compensatory response to the bottom-up noise driving delusion formation.

**A dynamic model of psychotic symptom emergence**

We present a model that aims to explain the development of psychotic symptoms in the context of a dynamic neural system evolved to maintain allostasis. Our model outlines a pathophysiological process that begins with NMDA receptor hypofunction (48) leading to heightened cortical excitability (49,50) and bottom-up noise (51). We propose that this initial disruption leads to delusional belief formation through aberrant prediction error signaling (52,53). We argue that this noisy input drives learning about the unreliability of new sensory information (54,55), resulting in an adaptive reliance on prior beliefs, delusion formation, hallucination propensity (12,40,41,56,57), and solidification of existing beliefs consistent with conversion to psychosis. The following sections outline the evidence for each of these steps and consider the predictions they make in studies examining factors leading to the emergence of psychotic symptoms.

*NMDA hypofunction leads to cortical hyperexcitability*

Studies conducted between the 1960s and the 1990s revealed that N-methyl-D-aspartate (NMDA) glutamate receptor antagonists such as phencyclidine (PCP) and ketamine can induce psychotic-like symptoms, leading to the glutamate hypothesis of schizophrenia (58–60). A growing body of evidence now supports the involvement of the excitatory neurotransmitter glutamate in schizophrenia and emphasizes a central role for NMDA receptors (61,62). Recent genomic discoveries implicating myriad NMDA-related genes further strengthen this theory (63,64).

Glutamatergic dysregulation appears to occur upstream of dopaminergic dysfunction and may act as a primary step in the pathophysiological process (65). Animal models have been instrumental in elucidating the consequences of NMDA receptor hypofunction on cortical activity and behavior (66). Electrophysiological recordings show that NMDA receptors preferentially regulate the firing rates of fast-spiking gamma-aminobutyric acid (GABA) interneurons, responsible for controlling the activity of glutamatergic pyramidal neurons via inhibitory feedback loops (67). It has been proposed that reduced NMDA receptor activity therefore results in disinhibition of pyramidal neurons, increasing their firing rate (68). This is exacerbated by allostatic downregulation of GABAergic signaling due to reduced excitatory drive to interneurons (69). Non-glutamatergic and GABAergic alterations associated with schizophrenia may further contribute to heightened excitability (70,71), as could environmental predictors of psychosis risk like stress (72).

*Cortical hyper-excitability is related to enhanced prediction-error signaling and leads to delusion formation.*

Hyper-excitability of cortex may be thought of in terms of aberrant prediction-error signaling: noisy input ascending the sensory processing hierarchy and leading to increased cortical firing corresponds to sensory 'information' that is not predicted by higher levels (51), causing inappropriate learning (52) by Hebbian principles of associative learning (73). In addition to ascending thalamocortical input disruptions, this may manifest as cortico-cortical dysconnectivity, as has been documented in psychosis (74). As reviewed above, a tendency toward delusion formation has been repeatedly linked to inappropriate learning via aberrant prediction error signaling (33–36,75). Large-scale connectivity data also support that this state of affairs is prominent in the earliest phases of psychosis, with enhanced corticothalamic connectivity evident in the Clinical High Risk state for Psychosis (CHR-P) and delusion-inducing psychotomimetic administration (76,77). At the level of conscious perception, these aberrant prediction errors may manifest as the well-documented, low-level sensory (and motor) dysfunction that has been observed early in psychosis (78).

*Cortical hyper-excitability also leads to decreased signal-to-noise ratio (SNR) of incoming sensory information*

Ascending sensory noise results in a reduction in the signal-to-noise ratio of incoming information (79). Noisiness of neuronal firing within sensory processing streams diminishes both signal transduction and discrimination performance (80,81), and may be most marked in the auditory system (82,83).

Internal sensory noise can also be estimated through the measurement of absolute detection threshold and as degree of deviation from Weber's law at near-threshold stimulus intensities on psychophysical tasks (84). Decades of work has established pan-modal deficits in detection and discrimination abilities in schizophrenia (85,86), with especially notable deficits in the auditory domain by some measures (87). These include electrophysiological signatures like sensory gating failures that could be explained by heightened cortical excitability and resulting internal noise increases (88,89). This noise has also been linked to higher-level disruptions in working memory encoding that are characteristic of schizophrenia (88).

*Decreased SNR leads to adaptive prior hyper-precision*

An adaptive response to low SNR is slower learning (90) and more relative reliance on prior knowledge during inference in the face of learned sensory unreliability (91). This process is analogous to the reliability weighting that occurs between sensory modalities during Bayesian integration: information from the senses is combined to form an integrated posterior shaped by the reliability of each channel of input (92,93). This reliability weighting explains multisensory illusions like the ventriloquist effect (94) and the rubber hand illusion (95). In a similar way, perceptual belief updating depends on the reliability of incoming sensory evidence relative to that of priors (96,97). As reviewed above, prior hyper-precision is consistently found in hallucination-prone individuals and seems to track with symptom severity. Recent evidence also highlights an additional potential role for reduced sensory precision in hallucination-prone individuals (43).

While delusion formation has been tied to inappropriate learning via aberrant prediction error signaling, as outlined above, delusional conviction appears to be tied to slowness in updating existing beliefs in the face of disconfirmatory evidence (98) and a relatively over-weighting of priors during inference (38,99). Increased conviction in the reality of these beliefs and percepts is diagnostic of a transition to full psychosis (100).

**Convergent evidence**

The model we outline has the potential to explain a number of clinical and epidemiological findings in psychosis, including mechanisms behind established risk factors for psychosis development and the presence of other hallucinatory syndromes. We outline these briefly below.

*Environmental risk factors for psychosis increase bottom-up noise*

If, as our model states, psychosis begins with aberrant bottom-up noise, factors that cause or mimic that bottom-up noise should increase risk for psychosis. Exposure to several bioactive compounds increases the likelihood of psychosis development (101), likely via different specific pathways depending on the specific compound, population, and diagnosis in question.

Cannabis exposure has been considered to meet some criteria for causation of psychotic disorders lasting beyond transiently-induced psychotic symptoms (102). Cannabis exposure has since been shown to carry one of the largest risk ratios of any known environmental exposure for psychosis development (103) and worsening of psychotic symptoms (104). High cannabis use has been related to the presence of odd beliefs, magical thinking, unusual perceptual experience, and odd and eccentric behavior, as well as deficits in attention, psychomotor speed, working memory, cognitive flexibility, visuo-spatial processing, and verbal memory (105). Electrophysiologically, cannabis exposure has been linked strongly to higher Lempel Ziv Complexity (LZC), a measure of ascending noise (106), which may also result in diminished ERPs to incoming stimuli (107).

Serotonergic psychedelics (SPs) may also increase risk for psychosis (108) and produce prominent psychotomimetic effects in healthy users (109). While the link between SPs and psychosis remains hotly debated, numerous case series and reports exist of psychotic episodes precipitated by psychedelics (110,111) and recent studies suggest subclinical psychotic-like perceptual and cognitive alterations following use (112). SPs also appear to increase cortical excitability and bottom-up noise. At the cellular level, SPs excite layer V neurons via 5-HT2A receptor agonism, causing Gq-mediated intracellular calcium release (113). Similar to cannabis above, SP administration invariably increases LZC in human EEG studies (114–116). SPs also consistently decrease alpha band power (116,117), sometimes with coincident increases in gamma power (116,118) — consistent with an increase in cortical excitability (119) and bottom-up activity (120), respectively. Also consistent, fMRI studies report that SPs increase the entropy of BOLD signal in the cortex (121,122) and diminish existing correlations within and anticorrelations between resting state networks (117,123,124)—both strongly suggesting increased noise in cortical activity. Even phenomenologically, SP-induced perceptual abnormalities are in line with increased bottom-up noise (125), consisting primarily of elementary features (eg. patterns, lattices, added motion, simple sounds) as compared to the complex prior-driven hallucinations of psychosis such as figures and speech (126). Finally, consistent with the notion that SPs increase bottom-up noise to push vulnerable individuals towards psychosis, patients in early stages of schizophrenia experience worsened symptoms/decompensation after SP administration (127). Compellingly, patients in later stages of psychosis, when we argue compensatory prior hyper-precision has occurred, require higher doses of SPs to achieve the same psychotomimetic effects (125).

*Other hallucinatory syndromes*

Not all hallucinations occur in the context of delusions. In our model, any factors that decrease SNR–not just those that augment bottom-up noise–should increase risk for hallucinogenesis. What's more, those that decrease SNR because of diminished signal should produce hallucinations largely in the absence of delusional ideation. Across all sensory modalities, there are hallucinatory syndromes that begin with disruption of modality-specific sensory signals and result in modality-specific hallucinations that are canonically unaccompanied by delusions.

In one canonical example, Charles-Bonnet Syndrome (CBS) is characterized by complex and recurrent visual hallucinations that occur in the absence of other neuropsychiatric symptoms and are most often associated with vision loss (128). CBS has been attributed to diminished sensory visual input, which increases excitability in visual association cortices (129). Electrophysiological studies support this theory, showing increased occipital activity during CBS-associated visual hallucinations, as well as strikingly elevated visual cortical responses to peripheral field stimulation in patients with age-related macular degeneration compared with patients without hallucinations and controls (130,131). While it is unclear whether this cortical hyperexcitability contributes to CBS development or occurs as a consequence of it, it likely decreases the SNR and induces a state of hyperprecise priors that could give rise to rich visual hallucinatory experiences.

These phenomena are conserved across sensory modalities. Phantom limb syndrome is a condition in which patients experience sensations in a limb that does not exist (132). It is most frequently experienced after the physical loss of a limb and typically has a chronic course, often resistant to treatment (133). Musical ear syndrome is an uncommon phenomenon described as the perception of auditory musical sensations not corresponding to any external stimulus (134). It occurs more frequently following hearing disruption/loss and is considered the auditory equivalent of Charles-Bonnet Syndrome (135). Overall, there is evidence that deprivation of structured sensory input across different modalities can induce hallucinatory experiences (136,137). Importantly, although some of these syndromes have been shown to increase excitability of modality-specific cortex, none of them are consistently paired with delusional ideation, which is driven by supramodal ascending noise that drives new learning in our model.

**Model Predictions**

Pathophysiological models accounting for symptom emergence and dynamics posit strong causal links between their elements. This means that they are falsifiable, generating specific hypotheses about relationships between findings across levels of description. We provide a selection of these hypotheses below.

*There should be a temporal order to the computational abnormalities linked to psychosis.*

Our model implies an order of events starting with molecular and algorithmic changes that precede any symptom onset and culminating in psychosis onset (**Fig. 1**). We expect early emergence of glutamatergic signaling abnormalities and increased bottom-up noise, both of which should be detectable via electrophysiological measures and generative modeling. This should predict the emergence of aberrant learning, which would then lead to the formation of odd beliefs. Ascending noise should drive increased beliefs in the volatility of the environment and subsequent paranoia emergence. On the molecular level, we would not expect dopaminergic abnormalities to be present until the compensatory switch toward increased prior precision, which should correspond to hallucination emergence and the increased delusional conviction that is diagnostic of conversion to psychosis.

*Delusion emergence should precede hallucination emergence in most with psychosis.*

The order of events outlined above should make strong predictions not only about computational measures but also about clinical observations. If hallucinogenesis and delusion formation were independent processes, the temporal order of symptom onset would be equivocal among patients who experience both symptoms. However, we (138) and others (139) have recently shown that delusions tend to precede hallucinations in the natural history of psychosis. Of note, these data also demonstrate that delusions re-emerge after hallucinations, and that there may also be an inverse relationship between the severity of these symptoms as they emerge, consistent with a compensatory account of hallucinogenesis.

*Medications that decrease the ratio of precision of priors to precision of sensory evidence should decrease hallucinations and delusional conviction but not alter bottom-up noise or aberrant learning*

Medications used to treat fully-formed delusions and hallucinations decrease the relative precision of priors relative to that of incoming sensory information. Dopamine regulates a gain-control mechanism leading to increased relative prior precision (40,56). Traditional antipsychotics, which primarily act via D2 receptor blockade, thus decrease that precision. By contrast, mounting evidence indicates that cholinergic tone increases the precision of incoming sensory evidence (140,141). This helps explain why cholinesterase inhibitors, which increase cholinergic tone, are highly effective in managing the visual hallucinations of dementia with Lewy bodies (142). They also reduce the tendency of people with Lewy body disease to perceive meaningful images in visual noise, as our model would predict (143). The first line of antipsychotics not based in dopamine blockade has recently emerged and is based in enhancing muscarinic (e.g., xanomeline and KarXT) and nicotinic signaling (e.g., encenicline) (144–146). While both classes of medications should be effective in decreasing hallucination severity and delusional intensity, we might predict them to be less immediately effective in decreasing inappropriate learning via aberrant prediction error signaling driven by ascending glutamatergic input.

*Precision of incoming sensory evidence should be controlled via an independent mechanism from precision of priors.*

Because aberrant PE signaling and new belief formation appears to persist into established psychosis, hyper-precise priors must only partially compensate for aberrant PE signaling. This implies the existence of an initial, fixed deficit in the form of increased ascending noise as outlined above, and a secondary compensatory change that varies over time and can be most immediately linked to symptom expression (13). Although the temporal order of emergence is our focus here, it is worth noting that the subsequent illness course is characterized by a shifting dynamic balance between these two factors, despite what may be a state of equilibrium after emergence of frank psychosis (138). This means that physiological changes that increase cortical excitability (e.g., cannabis (see above), trauma exposure (147), or the hormonal shifts attendant to menopause and the postpartum period (148)) may lead to a worsening of hallucination severity via dynamic compensation. It also means that environmental factors and therapeutic interventions that influence compensatory prior precision could alter severity most immediately (149,150).

## Conclusions

It is worth stating explicitly that the model we propose is not complete as outlined, nor is it intended to be. Although we focus on a progression of illness that could explain the disease progression of a substantive proportion of individuals with psychosis, we also recognize the high value in finding the exceptions: knowledge of distinct pathophysiological trajectories is the start of individualized treatment and prevention. In this spirit, we hope that the framework we propose serves as an entry point for considering how information processing abnormalities underlying psychotic symptoms might be linked across time to result in disease expression, regardless of the individual routes taken toward illness.

Because our model includes causal relationships between factors that may be directly observed, measured, or estimated using computational modeling, its main elements should be falsifiable. Future work testing these relationships will be extremely valuable in refuting them or elaborating a more nuanced model of psychosis evolution over time within a Bayesian framework. Regardless of whether individual elements of our model survive the scrutiny of time, we hope that these ideas will spur the field to regard a deeper understanding of psychotic symptom development as a critical step in identifying the dynamic mechanisms ultimately giving rise to disease. Recognizing the evolution of symptoms from unobserved states will allow the extraordinary heterogeneity of psychosis to be parsed into pathophysiologically distinct subgroups that may be intervened upon for personalized treatment and prevention.

**Figure legend**

**Figure 1. A model of psychotic symptom development.** We propose a model linking molecular abnormalities to information processing abnormalities across levels of analysis and describe how these elements might be causally linked. We begin with NMDA hypofunction, which drives ascending noise, inappropriate learning via aberrant prediction error signaling, and delusion formation. This ascending noise also decreases the signal-to-noise ratio (SNR) of incoming evidence, which then leads to an adaptive down-weighting of that evidence relative to prior beliefs, which become relatively hyper-precise. This compensatory shift then leads to hallucinogenesis, crystallization of delusional beliefs, and frank psychosis onset.

**Acknowledgments**

We would like to expression deep appreciation to those who have helped us form these ideas and develop them into this manuscript, including Drs. Tyrone Cannon, Scott Woods, and Phillip Sterzer.

**Financial Disclosures**

ARP was supported by a K23 Career Development Award, one R21, and two R01s from the National Institute of Mental Health (K23MH115252-01A1; 5R21MH122940-02; R01MH129721; R01MH131768), by a Career Award for Medical Scientists from the Burroughs-Wellcome Fund, a Carol and Eugene Ludwig Award for Early Career Research, and by the Yale Department of Psychiatry and the Yale School of Medicine. All other authors report no biomedical financial interests or potential conflicts of interest.

**References**

1. Thorwald J (1957): *The Century of the Surgeon*. New York : Pantheon.

2. Bliss M (2011): *The Making of Modern Medicine: Turning Points in the Treatment of Disease.* University of Chicago Press.

3. Insel TR, Quirion R (2005): Psychiatry as a clinical neuroscience discipline. *JAMA* 294: 2221–2224.

4. Cupo L, McIlwaine SV, Daneault J-G, Malla AK, Iyer SN, Joober R, Shah JL (2021): Timing, Distribution, and Relationship Between Nonpsychotic and Subthreshold Psychotic Symptoms Prior to Emergence of a First Episode of Psychosis. *Schizophr Bull* 47: 604–614.

5. Browning M, Carter C, Chatham C, den Ouden H, Gillan C, Baker J, *et al.* (n.d.): Realizing the Clinical Potential of Computational Psychiatry: Report from the Banbury Center Meeting, February 2019.

6. Friston K (2010): The free-energy principle: a unified brain theory? *Nat Rev Neurosci* 11: 127–138.

7. Wang XJ, Krystal JH (2014): Computational psychiatry. *Neuron* 84: 638–654.

8. Murray JD, Demirtaş M, Anticevic A (2018): Biophysical Modeling of Large-Scale Brain Dynamics and Applications for Computational Psychiatry. *Biol Psychiatry Cogn Neurosci Neuroimaging* 3: 777–787.

9. Frässle S, Lomakina EI, Razi A, Friston KJ, Buhmann JM, Stephan KE (2017): Regression DCM for fMRI. *Neuroimage* 155: 406–421.

10. Wise T, Robinson OJ, Gillan CM (2023): Identifying Transdiagnostic Mechanisms in Mental Health Using Computational Factor Modeling. *Biol Psychiatry* 93: 690–703.

11. Petzschner FH, Garfinkel SN, Paulus MP, Koch C, Khalsa SS (2021): Computational Models of Interoception and Body Regulation. *Trends Neurosci* 44: 63–76.

12. Powers AR, Mathys C, Corlett PR (2017): Pavlovian conditioning-induced hallucinations result from overweighting of perceptual priors. *Science* 357: 596–600.

13. Kafadar E, Fisher VL, Quagan B, Hammer A, Jaeger H, Mourgues C, *et al.* (2022): Conditioned Hallucinations and Prior Overweighting Are State-Sensitive Markers of Hallucination Susceptibility. *Biol Psychiatry* 92: 772–780.

14. Smith R, Feinstein JS, Kuplicki R, Forthman KL, Stewart JL, Paulus MP, *et al.* (2021): Perceptual insensitivity to the modulation of interoceptive signals in depression, anxiety, and substance use disorders. *Sci Rep* 11: 2108.

15. Verdonk C, Teed AR, White EJ, Ren X, Stewart JL, Paulus MP, Khalsa SS (2024): Heartbeat-evoked neural response abnormalities in generalized anxiety disorder during peripheral adrenergic stimulation. *Neuropsychopharmacology*. https://doi.org/10.1038/s41386-024-01806-5

16. Reiter AMF, Heinze H-J, Schlagenhauf F, Deserno L (2017): Impaired Flexible Reward-

Based Decision-Making in Binge Eating Disorder: Evidence from Computational Modeling and Functional Neuroimaging. *Neuropsychopharmacology* 42: 628–637.

17. American Psychiatric Association (2021): *Diagnostic and Statistical Manual of Mental Disorders (DSM-5)*. American Psychiatric Publishing.

18. Powers AR III, Kelley M, Corlett PR (2016): Hallucinations as top-down effects on perception. *Biol Psychiatry Cogn Neurosci Neuroimaging* 1: 393–400.

19. Sheldon AD, Kafadar E, Fisher V, Greenwald MS, Aitken F, Negreira AM, *et al.* (2022): Perceptual pathways to hallucinogenesis. *Schizophr Res* 245: 77–89.

20. Adams RA, Stephan KE, Brown HR, Frith CD, Friston KJ (2013): The Computational Anatomy of Psychosis. *Frontiers in Psychiatry*, vol. 4. https://doi.org/10.3389/fpsyt.2013.00047

21. Friston KJ (2005): Hallucinations and perceptual inference. *Behavioral and Brain Sciences*, vol. 28. pp 764–766.

22. Teufel C, Subramaniam N, Fletcher PC (2013): The role of priors in Bayesian models of perception. *Frontiers in Computational Neuroscience*, vol. 7. https://doi.org/10.3389/fncom.2013.00025

23. Hugdahl K (2009): "Hearing voices": auditory hallucinations as failure of top-down control of bottom-up perceptual processes. *Scand J Psychol* 50: 553–560.

24. Yu AJ, Dayan P (2005): Uncertainty, neuromodulation, and attention. *Neuron* 46: 681–692.

25. Knill DC, Pouget A (2004): The Bayesian brain: the role of uncertainty in neural coding and computation. *Trends Neurosci* 27: 712–719.

26. Sterzer P, Adams RA, Fletcher P, Frith C, Lawrie SM, Muckli L, *et al.* (2018): The Predictive Coding Account of Psychosis. *Biol Psychiatry* 84: 634–643.

27. Corlett PR, Taylor JR, Wang XJ, Fletcher PC, Krystal JH (2010): Toward a neurobiology of delusions. *Prog Neurobiol* 92: 345–369.

28. Corlett PR, Horga G, Fletcher PC, Alderson-Day B, Schmack K, Powers AR (2019): Hallucinations and Strong Priors. *Trends Cogn Sci* 23. https://doi.org/10.1016/j.tics.2018.12.001

29. Linson A, Parr T, Friston KJ (2019): Active inference, stressors, and psychological trauma: A neuroethological model of (mal)adaptive explore-exploit dynamics in ecological context. *Behav Brain Res* 380: 112421.

30. Pezzulo G, Parr T, Friston K (2024): Active inference as a theory of sentient behavior. *Biol Psychol* 108741.

31. Parr T, Pezzulo G, Friston KJ (2022): *Active Inference: The Free Energy Principle in Mind, Brain, and Behavior*. MIT Press.

32. O'Reilly JX, Jbabdi S, Behrens TEJ (2012): How can a Bayesian approach inform neuroscience? *Eur J Neurosci* 35: 1169–1179.

33. Corlett PR, Murray GK, Honey GD, Aitken MR, Shanks DR, Robbins TW, *et al.* (2007): Disrupted prediction-error signal in psychosis: evidence for an associative account of delusions. *Brain* 130: 2387–2400.

34. Corlett PR, Fletcher PC (2012): The neurobiology of schizotypy: fronto-striatal prediction error signal correlates with delusion-like beliefs in healthy people. *Neuropsychologia* 50: 3612–3620.

35. Stuke H, Weilnhammer VA, Sterzer P, Schmack K (2019): Delusion Proneness is Linked to a Reduced Usage of Prior Beliefs in Perceptual Decisions. *Schizophr Bull* 45: 80–86.

36. Corlett PR, Honey GD, Fletcher PC (2007): From prediction error to psychosis: ketamine as a pharmacological model of delusions. *J Psychopharmacol* 21: 238–252.

37. Corlett PR, Honey GD, Krystal JH, Fletcher PC (2011): Glutamatergic model psychoses: prediction error, learning, and inference. *Neuropsychopharmacology* 36: 294–315.

38. Sheffield JM, Smith R, Suthaharan P, Leptourgos P, Corlett PR (2023): Relationships between cognitive biases, decision-making, and delusions. *Sci Rep* 13: 9485.

39. Reed EJ, Uddenberg S, Suthaharan P, Mathys CD, Taylor JR, Groman SM, Corlett PR (2020): Paranoia as a deficit in non-social belief updating. *Elife* 9. https://doi.org/10.7554/eLife.56345

40. Cassidy CM, Balsam PD, Weinstein JJ, Rosengard RJ, Slifstein M, Daw ND, *et al.* (2018): A Perceptual Inference Mechanism for Hallucinations Linked to Striatal Dopamine. *Curr Biol* 28: 503–514 e4.

41. Alderson-Day B, Lima CF, Evans S, Krishnan S, Shanmugalingam P, Fernyhough C, Scott SK (2017): Distinct processing of ambiguous speech in people with non-clinical auditory verbal hallucinations. *Brain* 140: 2475–2489.

42. Zarkali A, Adams RA, Psarras S, Leyland L-A, Rees G, Weil RS (2019): Increased weighting on prior knowledge in Lewy body-associated visual hallucinations. *Brain Commun* 1: fcz007.

43. Benrimoh D, Fisher VL, Seabury R, Sibarium E, Mourgues C, Chen D, Powers A (2024): Evidence for Reduced Sensory Precision and Increased Reliance on Priors in Hallucination-Prone Individuals in a General Population Sample. *Schizophr Bull* 50: 349–362.

44. Teufel C, Subramaniam N, Dobler V, Perez J, Finnemann J, Mehta PR, *et al.* (2015): Shift toward prior knowledge confers a perceptual advantage in early psychosis and psychosis-prone healthy individuals. *Proc Natl Acad Sci U S A*. https://doi.org/10.1073/pnas.1503916112

45. Dzafic I, Larsen KM, Darke H, Pertile H, Carter O, Sundram S, Garrido MI (2021): Stronger Top-Down and Weaker Bottom-Up Frontotemporal Connections During Sensory Learning Are Associated With Severity of Psychotic Phenomena. *Schizophr Bull* 47: 1039–1047.

46. Charlton CE, Lepock JR, Hauke DJ, Mizrahi R, Kiang M, Diaconescu AO (2022): Atypical prediction error learning is associated with prodromal symptoms in individuals at clinical high risk for psychosis. *Schizophrenia (Heidelb)* 8: 105.

47. Hauke DJ, Charlton CE, Schmidt A, Griffiths JD, Woods SW, Ford JM, *et al.* (2023): Aberrant Hierarchical Prediction Errors Are Associated With Transition to Psychosis: A Computational Single-Trial Analysis of the Mismatch Negativity. *Biol Psychiatry Cogn Neurosci Neuroimaging* 8: 1176–1185.

48. Quiñones GM, Mayeli A, Yushmanov VE, Hetherington HP, Ferrarelli F (2021): Reduced GABA/glutamate in the thalamus of individuals at clinical high risk for psychosis. *Neuropsychopharmacology* 46: 1133–1139.

49. Gawande DY, S Narasimhan KK, Shelkar GP, Pavuluri R, Stessman HAF, Dravid SM (2023): GluN2D Subunit in Parvalbumin Interneurons Regulates Prefrontal Cortex Feedforward Inhibitory Circuit and Molecular Networks Relevant to Schizophrenia. *Biol Psychiatry* 94: 297–309.

50. Kehrer C, Maziashvili N, Dugladze T, Gloveli T (2008): Altered Excitatory-Inhibitory Balance in the NMDA-Hypofunction Model of Schizophrenia. *Front Mol Neurosci* 1: 6.

51. Wood ET, Cummings KK, Jung J, Patterson G, Okada N, Guo J, *et al.* (2021): Sensory over-responsivity is related to GABAergic inhibition in thalamocortical circuits. *Transl Psychiatry* 11: 39.

52. Adams RA (2019): Cortical Disinhibition, Attractor Dynamics, and Belief Updating in Schizophrenia. In: Cutsuridis V, editor. *Multiscale Models of Brain Disorders*. Cham: Springer International Publishing, pp 81–89.

53. Isomura T, Kotani K, Jimbo Y, Friston KJ (2023): Experimental validation of the free-energy principle with in vitro neural networks. *Nat Commun* 14: 4547.

54. Thuné H, Recasens M, Uhlhaas PJ (2016): The 40-Hz Auditory Steady-State Response in Patients With Schizophrenia: A Meta-analysis. *JAMA Psychiatry* 73: 1145–1153.

55. Javitt DC, Sweet RA (2015): Auditory dysfunction in schizophrenia: integrating clinical and basic features. *Nat Rev Neurosci* 16: 535–550.

56. Schmack K, Bosc M, Ott T, Sturgill JF, Kepecs A (2021): Striatal dopamine mediates hallucination-like perception in mice. *Science* 372. https://doi.org/10.1126/science.abf4740

57. O'Callaghan C, Hall JM, Tomassini A, Muller AJ, Walpola IC, Moustafa AA, *et al.* (2017): Visual Hallucinations Are Characterized by Impaired Sensory Evidence Accumulation: Insights From Hierarchical Drift Diffusion Modeling in Parkinson's Disease. *Biol Psychiatry Cogn Neurosci Neuroimaging* 2: 680–688.

58. Javitt DC, Zukin SR (1991): Recent advances in the phencyclidine model of schizophrenia. *Am J Psychiatry* 148: 1301–1308.

59. Olney JW, Newcomer JW, Farber NB (1999): NMDA receptor hypofunction model of schizophrenia. *J Psychiatr Res* 33: 523–533.

60. Javitt DC, Zukin SR, Heresco-Levy U, Umbricht D (2012): Has an angel shown the way? Etiological and therapeutic implications of the PCP/NMDA model of schizophrenia. *Schizophr Bull* 38: 958–966.

61. Hu W, MacDonald ML, Elswick DE, Sweet RA (2015): The glutamate hypothesis of schizophrenia: evidence from human brain tissue studies. *Ann N Y Acad Sci* 1338: 38–57.

62. Merritt K, McCutcheon RA, Aleman A, Ashley S, Beck K, Block W, *et al.* (2023): Variability and magnitude of brain glutamate levels in schizophrenia: a meta and mega-analysis. *Mol Psychiatry* 28: 2039–2048.

63. Singh T, Poterba T, Curtis D, Akil H, Al Eissa M, Barchas JD, *et al.* (2022): Rare coding variants in ten genes confer substantial risk for schizophrenia. *Nature* 604: 509–516.

64. Trubetskoy V, Pardiñas AF, Qi T, Panagiotaropoulou G, Awasthi S, Bigdeli TB, *et al.* (2022): Mapping genomic loci implicates genes and synaptic biology in schizophrenia. *Nature* 604: 502–508.

65. McCutcheon RA, Krystal JH, Howes OD (2020): Dopamine and glutamate in schizophrenia: biology, symptoms and treatment. *World Psychiatry* 19: 15–33.

66. Nakao K, Singh M, Sapkota K, Fitzgerald A, Hablitz JJ, Nakazawa K (2022): 5-HT2A receptor dysregulation in a schizophrenia relevant mouse model of NMDA receptor hypofunction. *Transl Psychiatry* 12: 168.

67. Homayoun H, Moghaddam B (2007): NMDA receptor hypofunction produces opposite effects on prefrontal cortex interneurons and pyramidal neurons. *J Neurosci* 27: 11496–11500.

68. Moghaddam B, Javitt D (2012): From revolution to evolution: the glutamate hypothesis of schizophrenia and its implication for treatment. *Neuropsychopharmacology* 37: 4–15.

69. Krystal JH, Anticevic A, Yang GJ, Dragoi G, Driesen NR, Wang X-J, Murray JD (2017): Impaired Tuning of Neural Ensembles and the Pathophysiology of Schizophrenia: A Translational and Computational Neuroscience Perspective. *Biol Psychiatry* 81: 874–885.

70. Hommersom MP, Doorn N, Puvogel S, Lewerissa EI, Mordelt A, Ciptasari U, *et al.* (2024): CACNA1A haploinsufficiency leads to reduced synaptic function and increased intrinsic excitability. *bioRxiv*. https://doi.org/10.1101/2024.03.18.585506

71. Damaj L, Lupien-Meilleur A, Lortie A, Riou É, Ospina LH, Gagnon L, *et al.* (2015): CACNA1A haploinsufficiency causes cognitive impairment, autism and epileptic encephalopathy with mild cerebellar symptoms. *Eur J Hum Genet* 23: 1505–1512.

72. Dubé CM, Molet J, Singh-Taylor A, Ivy A, Maras PM, Baram TZ (2015): Hyper-excitability and epilepsy generated by chronic early-life stress. *Neurobiol Stress* 2: 10–19.

73. Hebb D (1949): The organization of behavior. EmphNew york. Wiley.

74. Crossley NA, Mechelli A, Fusar-Poli P, Broome MR, Matthiasson P, Johns LC, *et al.* (2009): Superior temporal lobe dysfunction and frontotemporal dysconnectivity in subjects at risk of psychosis and in first-episode psychosis. *Hum Brain Mapp* 30: 4129–4137.

75. Corlett PR, Honey GD, Aitken MR, Dickinson A, Shanks DR, Absalom AR, *et al.* (2006): Frontal responses during learning predict vulnerability to the psychotogenic effects of ketamine: linking cognition, brain activity, and psychosis. *Arch Gen Psychiatry* 63: 611–621.

76. Anticevic A, Corlett PR, Cole MW, Savic A, Gancsos M, Tang Y, *et al.* (2014): N-Methyl-D-Aspartate Receptor Antagonist Effects on Prefrontal Cortical Connectivity Better Model Early Than Chronic Schizophrenia. *Biol Psychiatry*. https://doi.org/10.1016/j.biopsych.2014.07.022

77. Anticevic A, Haut K, Murray JD, Repovs G, Yang GJ, Diehl C, *et al.* (2015): Association of Thalamic Dysconnectivity and Conversion to Psychosis in Youth and Young Adults at Elevated Clinical Risk. *JAMA Psychiatry* 72: 882–891.

78. Cutting J, Dunne F (1986): The Nature of the Abnormal Perceptual Experiences at the Onset of Schizophrenia. *Psychopathology* 19: 347–352.

79. Marks L (2014): *Sensory Processes: The New Psychophysics*. Elsevier.

80. Tolhurst DJ, Movshon JA, Dean AF (1983): The statistical reliability of signals in single neurons in cat and monkey visual cortex. *Vision Res* 23: 775–785.

81. Barlow HB (1956): Retinal noise and absolute threshold. *J Opt Soc Am* 46: 634–639.

82. Spiegel MF, Green DM (1981): Two procedures for estimating internal noise. *J Acoust Soc Am* 70: 69–73.

83. Neri P (2010): How inherently noisy is human sensory processing? *Psychon Bull Rev* 17: 802–808.

84. Gescheider GA (2013): *Psychophysics: The Fundamentals*. Psychology Press.

85. Javitt DC, Freedman R (2015): Sensory Processing Dysfunction in the Personal Experience and Neuronal Machinery of Schizophrenia. *Am J Psychiatry* 172: 17–31.

86. Bates TC (2005): The panmodal sensory imprecision hypothesis of schizophrenia: reduced auditory precision in schizotypy. *Pers Individ Dif* 38: 437–449.

87. Carroll CA, Boggs J, O'Donnell BF, Shekhar A, Hetrick WP (2008): Temporal processing dysfunction in schizophrenia. *Brain Cogn* 67: 150–161.

88. Starc M, Murray JD, Santamauro N, Savic A, Diehl C, Cho YT, *et al.* (2017): Schizophrenia is associated with a pattern of spatial working memory deficits consistent with cortical disinhibition. *Schizophr Res* 181: 107–116.

89. Murray JD, Anticevic A, Gancsos M, Ichinose M, Corlett PR, Krystal JH, Wang XJ (2014): Linking microcircuit dysfunction to cognitive impairment: effects of disinhibition associated with schizophrenia in a cortical working memory model. *Cereb Cortex* 24: 859–872.

90. Schlimmer JC, Granger RH (1986): Incremental learning from noisy data. *Mach Learn* 1: 317–354.

91. Deneve S (2012): Making decisions with unknown sensory reliability. *Front Neurosci* 6: 75.

92. Ernst MO, Banks MS (2002): Humans integrate visual and haptic information in a statistically optimal fashion. *Nature* 415: 429–433.

93. Fetsch CR, Pouget A, DeAngelis GC, Angelaki DE (2011): Neural correlates of reliability-based cue weighting during multisensory integration. *Nat Neurosci* 15: 146–154.

94. Alais D, Burr D (2004): The ventriloquist effect results from near-optimal bimodal integration. *Curr Biol* 14: 257–262.

95. Chancel M, Ehrsson HH (2023): Proprioceptive uncertainty promotes the rubber hand

illusion. *Cortex* 165: 70–85.

96. Standage D, Wang D-H, Heitz RP, Simen P (2016): *Toward a Unified View of the Speed-Accuracy Trade-Off: Behaviour, Neurophysiology and Modelling*. Frontiers Media SA.

97. Kersten D, Mamassian P, Yuille A (2004): Object perception as Bayesian inference. *Annu Rev Psychol* 55: 271–304.

98. Penzenstadler L, Chatton A, Lecomte T, Huguelet P, Lecardeur L, Azoulay S, *et al.* (2020): Does the Beck Cognitive Insight Scale predict change in delusional beliefs? *Psychol Psychother* 93: 690–704.

99. Gawęda Ł, Staszkiewicz M, Balzan RP (2017): The relationship between cognitive biases and psychological dimensions of delusions: The importance of jumping to conclusions. *J Behav Ther Exp Psychiatry* 56: 51–56.

100. McGlashan TH, Walsh B, Woods S (2010): *The Psychosis-Risk Syndrome : Handbook for Diagnosis and Follow-Up*. New York: Oxford University Press, p xii, 243 p.

101. Pries L-K, Erzin G, van Os J, Ten Have M, de Graaf R, van Dorsselaer S, *et al.* (2021): Predictive Performance of Exposome Score for Schizophrenia in the General Population. *Schizophr Bull* 47: 277–283.

102. D'Souza DC, DiForti M, Ganesh S, George TP, Hall W, Hjorthøj C, *et al.* (2022): Consensus paper of the WFSBP task force on cannabis, cannabinoids and psychosis. *World J Biol Psychiatry* 23: 719–742.

103. Hjorthøj C, Posselt CM, Nordentoft M (2021): Development Over Time of the Population-Attributable Risk Fraction for Cannabis Use Disorder in Schizophrenia in Denmark. *JAMA Psychiatry* 78: 1013–1019.

104. Solmi M, De Toffol M, Kim JY, Choi MJ, Stubbs B, Thompson T, *et al.* (2023): Balancing risks and benefits of cannabis use: umbrella review of meta-analyses of randomised controlled trials and observational studies. *BMJ* 382: e072348.

105. D'Souza DC, Ganesh S, Cortes-Briones J, Campbell MH, Emmanuel MK (2020): Characterizing psychosis-relevant phenomena and cognitive function in a unique population with isolated, chronic and very heavy cannabis exposure. *Psychol Med* 50: 2452–2459.

106. Cortes-Briones JA, Cahill JD, Skosnik PD, Mathalon DH, Williams A, Sewell RA, *et al.* (2015): The psychosis-like effects of Delta(9)-tetrahydrocannabinol are associated with increased cortical noise in healthy humans. *Biol Psychiatry* 78: 805–813.

107. D'Souza DC, Fridberg DJ, Skosnik PD, Williams A, Roach B, Singh N, *et al.* (2012): Dose-related modulation of event-related potentials to novel and target stimuli by intravenous $\Delta^9$-THC in humans. *Neuropsychopharmacology* 37: 1632–1646.

108. Fiorentini A, Cantù F, Crisanti C, Cereda G, Oldani L, Brambilla P (2021): Substance-Induced Psychoses: An Updated Literature Review. *Front Psychiatry* 12: 694863.

109. Preller KH, Vollenweider FX (2018): Phenomenology, Structure, and Dynamic of Psychedelic States. *Curr Top Behav Neurosci* 36: 221–256.

110. Dos Santos RG, Bouso JC, Hallak JEC (2017): Ayahuasca, dimethyltryptamine, and psychosis: a systematic review of human studies. *Ther Adv Psychopharmacol* 7: 141–157.

111. Bowers MB Jr (1977): Psychoses precipitated by psychotomimetic drugs. A follow-up study. *Arch Gen Psychiatry* 34: 832–835.

112. Lebedev AV, Acar K, Horntvedt O, Cabrera AE, Simonsson O, Osika W, *et al.* (2023): Alternative beliefs in psychedelic drug users. *Sci Rep* 13: 16432.

113. Kwan AC, Olson DE, Preller KH, Roth BL (2022): The neural basis of psychedelic action. *Nat Neurosci* 25: 1407–1419.

114. Schartner MM, Carhart-Harris RL, Barrett AB, Seth AK, Muthukumaraswamy SD (2017): Increased spontaneous MEG signal diversity for psychoactive doses of ketamine, LSD and psilocybin. *Sci Rep* 7: 46421.

115. Mediano PAM, Rosas FE, Timmermann C, Roseman L, Nutt DJ, Feilding A, *et al.* (2024): Effects of External Stimulation on Psychedelic State Neurodynamics. *ACS Chem Neurosci*. https://doi.org/10.1021/acschemneuro.3c00289

116. Timmermann C, Roseman L, Schartner M, Milliere R, Williams LTJ, Erritzoe D, *et al.* (2019): Neural correlates of the DMT experience assessed with multivariate EEG. *Sci Rep* 9: 16324.

117. Carhart-Harris RL, Muthukumaraswamy S, Roseman L, Kaelen M, Droog W, Murphy K, *et al.* (2016): Neural correlates of the LSD experience revealed by multimodal neuroimaging. *Proc Natl Acad Sci U S A* 113: 4853–4858.

118. Kometer M, Pokorny T, Seifritz E, Volleinweider FX (2015): Psilocybin-induced spiritual experiences and insightfulness are associated with synchronization of neuronal oscillations. *Psychopharmacology* 232: 3663–3676.

119. Jensen O, Mazaheri A (2010): Shaping functional architecture by oscillatory alpha activity: gating by inhibition. *Front Hum Neurosci* 4: 186.

120. Merker B (2013): Cortical gamma oscillations: the functional key is activation, not cognition. *Neurosci Biobehav Rev* 37: 401–417.

121. Viol A, Palhano-Fontes F, Onias H, de Araujo DB, Viswanathan GM (2017): Shannon entropy of brain functional complex networks under the influence of the psychedelic Ayahuasca. *Sci Rep* 7: 7388.

122. Lebedev AV, Kaelen M, Lövdén M, Nilsson J, Feilding A, Nutt DJ, Carhart-Harris RL (09 2016): LSD-induced entropic brain activity predicts subsequent personality change. *Hum Brain Mapp* 37: 3203–3213.

123. Timmermann C, Roseman L, Haridas S, Rosas FE, Luan L, Kettner H, *et al.* (2023): Human brain effects of DMT assessed via EEG-fMRI. *Proc Natl Acad Sci U S A* 120: e2218949120.

124. Madsen MK, Stenbæk DS, Arvidsson A, Armand S, Marstrand-Joergensen MR, Johansen SS, *et al.* (2021): Psilocybin-induced changes in brain network integrity and segregation correlate with plasma psilocin level and psychedelic experience. *Eur Neuropsychopharmacol* 50: 121–132.

125. Suzuki K, Seth AK, Schwartzman DJ (2023): Modelling phenomenological differences in aetiologically distinct visual hallucinations using deep neural networks. *Front Hum Neurosci* 17: 1159821.

126. Leptourgos P, Fortier-Davy M, Carhart-Harris R, Corlett PR, Dupuis D, Halberstadt AL, *et al.* (2020): Hallucinations Under Psychedelics and in the Schizophrenia Spectrum: An Interdisciplinary and Multiscale Comparison. *Schizophr Bull* 46: 1396–1408.

127. Wolf G, Singh S, Blakolmer K, Lerer L, Lifschytz T, Heresco-Levy U, *et al.* (2023): Could psychedelic drugs have a role in the treatment of schizophrenia? Rationale and strategy for safe implementation. *Mol Psychiatry* 28: 44–58.

128. Kazui H, Ishii R, Yoshida T, Ikezawa K, Takaya M, Tokunaga H, *et al.* (2009): Neuroimaging studies in patients with Charles Bonnet Syndrome. *Psychogeriatrics* 9: 77–84.

129. Pang L (2016): Hallucinations Experienced by Visually Impaired: Charles Bonnet Syndrome. *Optom Vis Sci* 93: 1466–1478.

130. Painter DR, Dwyer MF, Kamke MR, Mattingley JB (2018): Stimulus-Driven Cortical Hyperexcitability in Individuals with Charles Bonnet Hallucinations. *Curr Biol* 28: 3475–3480.e3.

131. Piarulli A, Annen J, Kupers R, Laureys S, Martial C (2021): High-Density EEG in a Charles Bonnet Syndrome Patient during and without Visual Hallucinations: A Case-Report Study. *Cells* 10. https://doi.org/10.3390/cells10081991

132. Brugger P (2012): Phantom Limb, Phantom Body, Phantom Self: A Phenomenology of "Body Hallucinations." In: Blom JD, Sommer IEC, editors. *Hallucinations: Research and Practice*. New York, NY: Springer New York, pp 203–218.

133. Chahine L, Kanazi G (2007): Phantom limb syndrome: a review. *Middle East J Anaesthesiol* 19: 345–355.

134. Evers S (2006): Musical hallucinations. *Curr Psychiatry Rep* 8: 205–210.

135. Sanchez TG, Rocha SCM, Knobel KAB, Kii MA, Santos RMR dos, Pereira CB (2011): Musical hallucination associated with hearing loss. *Arq Neuropsiquiatr* 69: 395–400.

136. Mason OJ, Brady F (2009): The psychotomimetic effects of short-term sensory deprivation. *J Nerv Ment Dis* 197: 783–785.

137. Wackermann J, Pütz P, Allefeld C (2008): Ganzfeld-induced hallucinatory experience, its phenomenology and cerebral electrophysiology. *Cortex* 44: 1364–1378.

138. Mourgues-Codern C, Benrimoh D, Gandhi J, Farina EA, Vin R, Zamorano T, *et al.* (2024, February 20): Emergence and dynamics of delusions and hallucinations across stages in early psychosis. *arXiv [q-bio.NC]*. Retrieved from http://arxiv.org/abs/2402.13428

139. Hermans K, van der Steen Y, Kasanova Z, van Winkel R, Reininghaus U, Lataster T, *et al.* (2020): Temporal dynamics of suspiciousness and hallucinations in clinical high risk and first episode psychosis. *Psychiatry Res* 290: 113039.

140. Moran RJ, Campo P, Symmonds M, Stephan KE, Dolan RJ, Friston KJ (2013): Free

energy, precision and learning: the role of cholinergic neuromodulation. *J Neurosci* 33: 8227–8236.

141. Marshall L, Mathys C, Ruge D, de Berker AO, Dayan P, Stephan KE, Bestmann S (2016): Pharmacological Fingerprints of Contextual Uncertainty. *PLoS Biol* 14: e1002575.

142. Ukai K, Fujishiro H, Iritani S, Ozaki N (2015): Long-term efficacy of donepezil for relapse of visual hallucinations in patients with dementia with Lewy bodies. *Psychogeriatrics* 15: 133–137.

143. Yokoi K, Nishio Y, Uchiyama M, Shimomura T, Iizuka O, Mori E (2014): Hallucinators find meaning in noises: pareidolic illusions in dementia with Lewy bodies. *Neuropsychologia* 56: 245–254.

144. Sauder C, Allen LA, Baker E, Miller AC, Paul SM, Brannan SK (2022): Effectiveness of KarXT (xanomeline-trospium) for cognitive impairment in schizophrenia: post hoc analyses from a randomised, double-blind, placebo-controlled phase 2 study. *Transl Psychiatry* 12: 491.

145. Leber A, Ramachandra R, Ceban F, Kwan ATH, Rhee TG, Wu J, *et al.* (2024): Efficacy, safety, and tolerability of xanomeline for schizophrenia spectrum disorders: a systematic review. *Expert Opin Pharmacother* 1–10.

146. Keefe RSE, Meltzer HA, Dgetluck N, Gawryl M, Koenig G, Moebius HJ, *et al.* (2015): Randomized, Double-Blind, Placebo-Controlled Study of Encenicline, an α7 Nicotinic Acetylcholine Receptor Agonist, as a Treatment for Cognitive Impairment in Schizophrenia. *Neuropsychopharmacology* 40: 3053–3060.

147. Rossi S, De Capua A, Tavanti M, Calossi S, Polizzotto NR, Mantovani A, *et al.* (2009): Dysfunctions of cortical excitability in drug-naïve posttraumatic stress disorder patients. *Biol Psychiatry* 66: 54–61.

148. Manyukhina VO, Orekhova EV, Prokofyev AO, Obukhova TS, Stroganova TA (2022): Altered visual cortex excitability in premenstrual dysphoric disorder: Evidence from magnetoencephalographic gamma oscillations and perceptual suppression. *PLoS One* 17: e0279868.

149. Powers AR, Bien C, Corlett PR (2018): Aligning computational psychiatry with the hearing voices movement hearing their voices. *JAMA Psychiatry* 75. https://doi.org/10.1001/jamapsychiatry.2018.0509

150. Benrimoh D, Sheldon A, Sibarium E, Powers AR (2021): Computational Mechanism for the Effect of Psychosis Community Treatment: A Conceptual Review From Neurobiology to Social Interaction. *Front Psychiatry* 12: 685390.

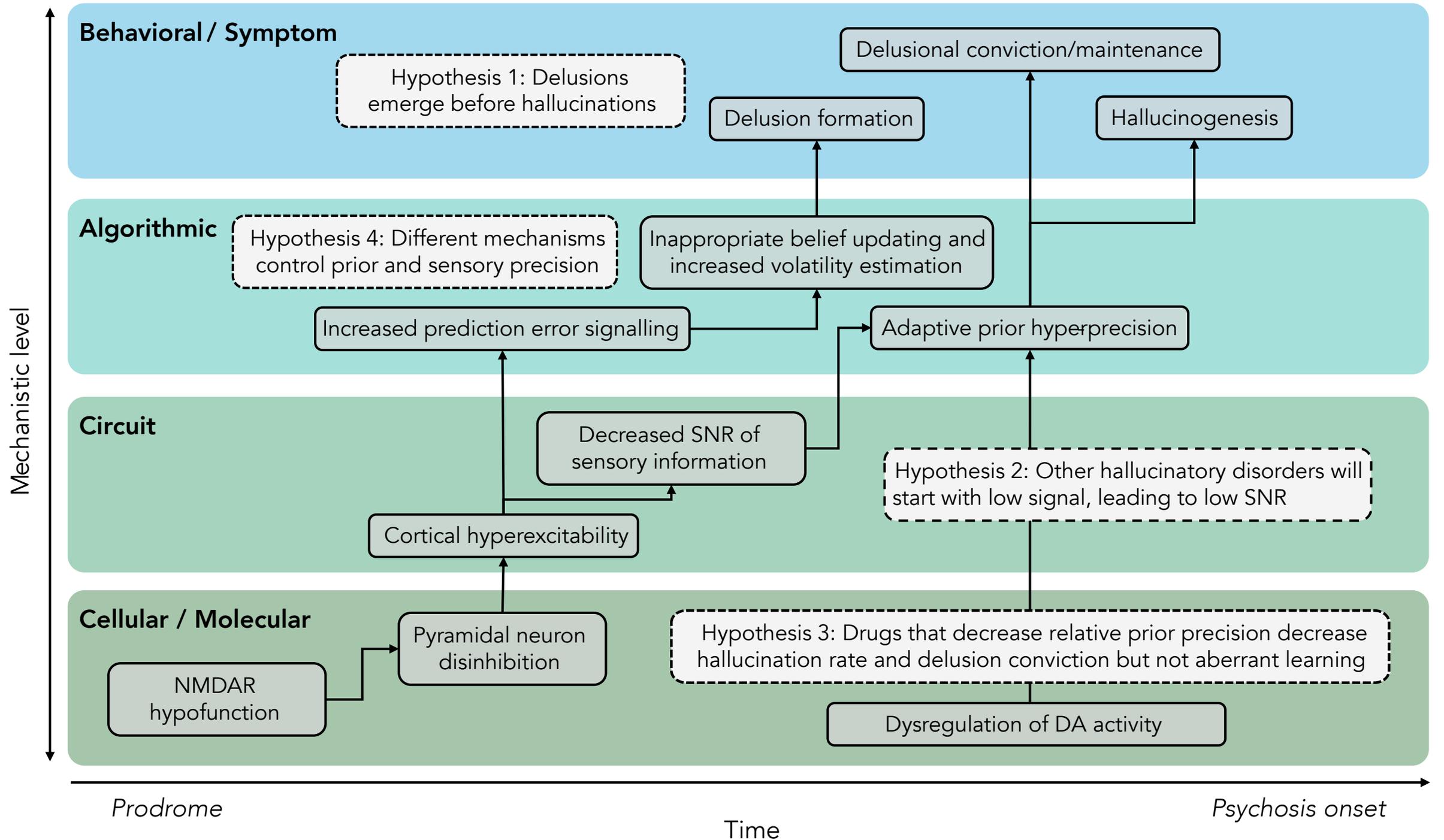